\documentclass[twocolumn,showpacs,showkeys,preprintnumbers,amsmath,amssymb]{revtex4} 
\usepackage{graphicx}% Include figure files
\usepackage{dcolumn}% Align table columns on decimal point
\usepackage{bm}% bold math

\begin{document}

\title{Efficient dynamical nuclear polarization in quantum dots: Temperature dependence}
% Force line breaks with \\

\author{B.\ Urbaszek$^1$}
\email[Corresponding author : ]{urbaszek@insa-toulouse.fr}
\author{P.-F.\ Braun$^1$}
\author{X.\ Marie$^1$}
\author{O.\ Krebs$^2$}
\author{A.\ Lemaitre$^2$}
\author{P.\ Voisin$^2$}
\author{T.\ Amand$^1$}

\affiliation{%
$^1$LPCNO, INSA-CNRS-UPS, 135 avenue de Rangueil, 31077 Toulouse Cedex 4, France}

\affiliation{%
$^2$Laboratoire de Photonique et Nanostructures, route de Nozay, 91460 Marcoussis, France}

\date{\today}

\begin{abstract}
We investigate in micro-photoluminescence experiments the dynamical nuclear polarization in individual InGaAs quantum dots. Experiments carried out in an applied magnetic field of 2T show that the nuclear polarization achieved through the optical pumping of electron spins is increasing with the sample temperature between 2K and 55K, reaching a maximum of about 50\%. Analysing the dependence of the Overhauser shift on the spin polarization of the optically injected electron as a function of temperature enables us to identify the main reasons for this increase.  
 
\end{abstract}

\pacs{72.25.Fe,73.21.La,78.55.Cr,78.67.Hc}% PACS, the Physics and Astronomy
                             % Classification Scheme
                             % 78.55.Cr   III-V semiconductors  
                             % 73.21.La    Quantum dots   
                             % 78.66.Hc  Optical properties 
                            \keywords{Quantum dots, hyperfine interaction}%Use showkeys class option if keyword
                             %display desired
\maketitle

\textbf{Introduction}

The spin of a single carrier in a semiconductor quantum dot (QD) is accessible via the optical selection rules and several groups have shown that it is possible to write, manipulate and read out electron spin states in QDs in ultra-fast optical experiments \cite{Cortez02,dutt,greilich,AwschScience}. Although the electron spin in a QD is stable up to ms under certain conditions \cite{Kroutvar1}, it is eventually too short lived to store a preferred spin direction and ultimately a superposition of spin states. Nuclear spins of the atoms in the QD are better suited for this purpose, with spin lifetimes of several minutes measured in bulk GaAs \cite{paget82}. Two recent proposals that are motivated by the possibility to use the nuclear spins as a quantum memory aim to combine the strengths of electron-spin (or charge) manipulation with the long-term memory provided by the nuclear spin system \cite{taylor, christ}. The target is to  achieve a nuclear polarization degree that is high enough (close to $100\%$) to transfer a coherent spin state from a single electron to the nuclear spin system. The electron spin in an InGaAs QD strongly interacts with about $10^5$ nuclear spins via the hyperfine interaction.  Dynamical polarization of the nuclear spins can be achieved through the repeated optical injection of an electron with a well defined spin orientation that undergoes a simultaneous spin-flip with a nuclear spin (flip-flop)\cite{opor,GammonPRL}. To construct an effective nuclear field $B_n$ of the order of several Tesla in InGaAs QDs through optical pumping, the laser excitation power and the applied magnetic field have to be optimised \cite{braun2006,Imapriv,Tarta1}. The Zeeman splitting $\delta_z^e$ of an electron is given by the contributions of the external field $B_{ext}$ and $B_n$.  The contribution to $\delta_z^e$ due to $B_n$ is the Overhauser shift (OHS) $\delta_n$ and is a measure of the nuclear polarization degree. 
The application of $B_{ext}$ has to main effects: (i) It slows down certain nuclear relaxation processes \cite{braun2006} and (ii) it opens up an electron Zeeman splitting, that makes the spin flip-flop process at the origin of dynamical polarization energetically more difficult. 

\begin{figure}
\includegraphics[width=0.43\textwidth]{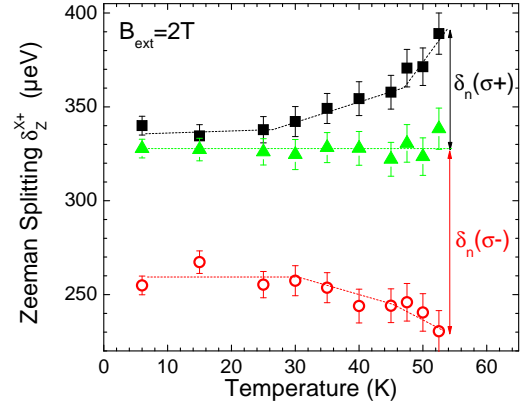}
\caption{\label{fig:fig1} Zeeman splitting $\delta_z^{X+}$ measured for the $X^+$ exciton PL of an individual QD for three different excitation laser polarizations: $\sigma^+$ (solid squares),$\sigma^-$ (hollow circles) and linear (solid triangles) in an applied magnetic field of 2T as a function of temperature. The dotted lines are a guide to the eye. The arrows show the OHS $\delta_n$ due to the effective nuclear field $B_n$, which is zero for linear excitation.
 }
\end{figure}

In this work we studied the temperature dependence of the dynamical nuclear polarization. We demonstrate that the sample temperature is a key parameter for maximising the nuclear polarization degree. Previous studies of the temperature dependence of hyperfine effects for weakly confined electrons, for example localised at donor sites, have been difficult to interpret as carriers get ionised (delocalised) \cite{opor,Dzhioev2}. The strong confinement in QDs provides the unique opportunity to study the physics of the coupled electron-nuclear spin system at high lattice temperature \cite{henne2007}. We find that the dynamical nuclear polarization for InGaAs QDs containing a singly positively charged exciton $X^+$ is more efficient at 55K than at 2K. This is very surprising, as both the electron and the nuclear spin lifetime decrease at higher lattice temperature when measured independently \cite{Marie01,LuNMR}. We show that raising the temperature makes the flip-flop process at the origin of the dynamical polarization less sensitive the energy separation $\delta_z^e$. Our measurements indicate that this is due to a level broadening of the spin up and spin down states due to the uncertainty principle, which allows the electron spin-flip to take place despite an energy mismatch.

\textbf{Experimental details}

The sample and the optical spectroscopy set-up are the same as in reference \cite{braun2006}. The sample contains the following layers, starting from the substrate: 200nm of p doped (Be) GaAs / 25nm of GaAs / InGaAs dots with a wetting layer grown in the Stranski-Krastanov mode / 30 nm of GaAs / 100nm Ga$_{0.7}$Al$_{0.3}$As / 20nm of GaAs. Placing a doped layer below the dots enables holes to tunnel into the dots. 
The photoluminescence (PL) measurements on individual dots were carried out with a confocal microscope connected to a spectrometer and a CCD camera giving a spectral precision of +/- 2.5 $\mu$eV. We excite directly the heavy hole to electron transition at 1.43 eV in the wetting layer with a pulsed Ti-Sapphire laser with an 80 MHz repetition frequency. 

We have done all the measurements on positively charged excitons X$^+$ containing an optically excited electron and a hole, with an additional hole originating from the Be doped layer. The circular polarization degree of the luminescence is defined as $P_c = (I^+ - I^-)/ (I^+ + I^-)$, where $I^{+(-)}$ is the $\sigma^{+(-)}$ polarized PL intensity.  Transitions with $P_c$ in the order of 50\% at zero external field are attributed to the X$^+$. For PL transitions stemming from neutral excitons in InGaAs dots $P_c$ is only in the order of a few percent due to anisotropic exchange \cite{Mb2,Senes2,Lang2004,Eble}. In this work we will focus on the dynamical polarization in InGaAs QDs in an external magnetic field $B_{ext}$ parallel to the sample growth direction, that is larger than both the local magnetic field $B_L$ (characterising the strength of the nuclear dipole-dipole interaction) and the Knight field $B_e$ (the effective magnetic field seen by the nuclei due to the presence of a spin polarized electron), which are in the order of mT\cite{Lai,Aki2006}. 

\textbf{Experimental results and discussion}
\begin{figure}
\includegraphics[width=0.42\textwidth]{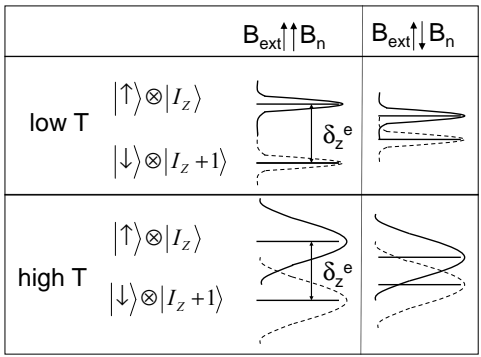}
\caption{\label{fig:fig2} Coupled electron-nuclear spin system with two states separated by an energy difference $\delta_z^e$. The levels are broadened by $\hbar/\tau_c$. The overlap between the two levels allows the dynamical nuclear polarization to build up via  the spin flip-flop process.
 }
\end{figure}
The presence of spin polarized electrons is essential for building up a nuclear field. This is the case during the radiative lifetime of the positively charged exciton X$^+$ (about 1 ns \cite{pif05}), where the holes form a spin singlet and the single electron interacts with the nuclei. The analysis of the circular polarization of the $X^+$ luminescence in QDs following circularly polarized laser excitation will thus probe \emph{directly} the spin polarization of the electron as $\langle\hat{S}_z^e\rangle=-P_c/2$.
We have measured the Zeeman splitting $\delta_z^{X+}$ of the $X^+$ exciton in an individual dot for sample temperatures up to 55K. The overall aim is to polarize as many nuclei as possible, the experiments were therefore carried out at an external field of 2 T, which gave a strong nuclear polarization in a previous study at T=2K \cite {braun2006}. 

The OHS $\delta_n$ for $\sigma^+$ and $\sigma^-$ excitation, respectively, is given by \cite {braun2006}: 

\begin{equation} 
\label{equOHS}
\delta_n(\sigma^{\pm})=\delta_Z^{X+}(\sigma^{\pm})-\delta_Z^{X+}(lin)=-g_e\mu_BB_n(\sigma^{\pm}) \\
\end{equation}

To relate $\delta_n$ to the nuclear polarization we take a dot composition of In$_{0.45}$Ga$_{0.55}$As \cite{AL2004} which corresponds to a maximum OHS of about 236$\mu eV$ for a nuclear polarization of 100\%. Below 30K, the nuclear polarization (OHS $\delta_n$) is much higher for $\sigma^-$ than for $\sigma^+$ excitation. As has been confirmed by several groups \cite{braun2006,Imapriv,Tarta1}, this asymmetry is due to the orientation of $B_n$ antiparallel to $B_{ext}$ in the case of $\sigma^-$ excitation, which minimises $\delta_z^e$. In contrast, for the $\sigma^+$ excitation $B_n$ and $B_{ext}$ are parallel (additive), resulting in a larger $\delta_z^e$, making the flip-flop process and therefore dynamical nuclear polarization less likely. A typical dot is shown in figure 1, but as a general rule, (i) we find a maximum $\delta_n$ for any dot at the highest temperature, for both excitation laser polarizations in the range 2K to 55K, with a maximum of about $\delta_n=110\mu eV$ that corresponds to a nuclear polarization of 47\%.
(ii) The ratio $\delta_n(\sigma^-):\delta_n(\sigma^+)$ decreases with temperatures, the graphs in figure 1 become more symmetric as the temperature is increased. This means that it gets easier for the electron to flip its spin despite a considerable Zeeman splitting. The fact that $B_n$ is added in one case to $B_{ext}$, and subtracted from $B_{ext}$ in the other case  does not play an important role at elevated temperatures. This can be understood if the energy levels involved in the flip-flop would get broader with temperature, as indicated in figure 2. The correlation time $\tau_c$ is the time an electron with a well defined spin does interact with the nuclear spins in the dot. One hypothesis is that $\tau_c$ decreases as the temperature is raised, resulting in a broadening of the electron Zeeman levels given by $\hbar/\tau_c$. This would increase the overlap of the energy levels and hence increase the rate of the flip-flop process, as indicated in figure 2. It should be noted that due to the small value of the nuclear gyromagnetic ratio the energy difference between the nuclear spins states is negligible compared to the electron Zeeman splitting $\delta_z^e$. 

\begin{figure}
\includegraphics[width=0.45\textwidth]{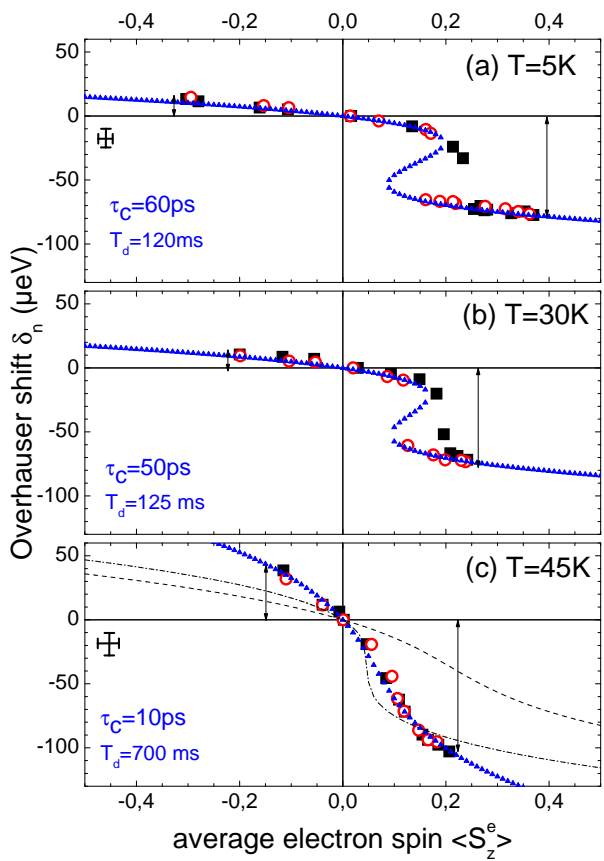}
\caption{\label{fig:fig3} The OHS $\delta_n$ is measured for the same dot as in figure 1 as a function of the optically injected electron spin at a constant magnetic field of 2 Tesla at a sample temperature of (a)5K, (b) 30K and (c)45K, respectively. The solid squares show the measurements when going from low to high electron spin polarization, the hollow circles from high to low. The bold lines in (a) to (c) are fits with equation \ref{eq:Model} using $\tilde{A}=48\mu eV$, $\tilde{Q}=13$, $N=60000$, $g_e=-0.48$, $f_e=0.05$ and the values of $\tau_c$ and $T_d$ indicated in the figure. The dashed and the dash-dotted line in (c) correspond to $\tau_c=10ps$ and $T_d=125ms$ and $\tau_c=50ps$ and $T_d=700ms$, respectively. The crosses in (a) and (c) indicate typical error bars.}
\end{figure}

One way of accessing $\tau_c$ is to fit the dependence of $\delta_n$ measured as a function of the average electron spin $\langle\hat{S}_z^e\rangle$, which we have done for different temperatures, see figure 3.
The nuclei get polarized by optically injected electrons with a preferred spin polarization. The average spin $\langle\hat{S}_z^e\rangle$ of the injected electron is varied by changing the relative retardation of the excitation beam continuously from $\lambda/4$ to $3\lambda/4$. Please note that as we record two spectra, $\sigma^+$ and $\sigma^-$ polarized, for any given excitation polarization, we measure both $\delta_n$ and the circular polarization degree $P_c$. As $\langle\hat{S}_z^e\rangle=-P_c/2$ we are able to measure $\delta_n$ as a function of the average electron spin $\langle\hat{S}_z^e\rangle$ in the dot \cite{Eble}.  

The hyperfine interaction between an electron of spin $\hat{S}^e=\frac{1}{2}\hat{\sigma}^e$ confined to a QD and $N$ nuclei is described by the Fermi contact Hamiltonian. 
\begin{equation}
\label{eq:eqHf}
\hat{H}_{hf} = \frac{\nu_0}{2}\sum_{j}A^j \vert\psi(\bar{r}_j)\vert^2 \left(2\hat{I}_z^j\hat{S}_z^e+ [\hat{I}_+^j\hat{S}_-^e+\hat{I}_-^j\hat{S}_+^e]\right)
\end{equation}
where $\nu_0$ is the two atom unit cell volume, $\bar{r}_j$ is the position of the nuclei $j$ with spin $\hat{I}^j$, the nuclear species are In, As and Ga. $A^j$ is the constant of the hyperfine interaction with the electron and $\psi(\bar{r})$ is the electron envelope function. Due to the $p$-symmetry of the periodic part of hole Bloch function the interaction of the hole via equation (2) is neglected in the following \cite{Abra}.

The model developped in reference \cite{Eble} gives an implicit expression for the OHS $\delta_n$ as a function of the correlation time of the hyperfine interaction $\tau_c$:

\begin{equation} 
\label{eq:Model}
\delta_n=2\tilde{A}\langle I_z \rangle= \frac{2\tilde{A}\tilde{Q}\langle \hat{S}_z^e \rangle}{1+\frac{T_{e1}(\delta_n)}{T_d}} 
\end{equation}

Where we have introduced $\tilde{A}$ as the average of the hyperfine constants $A^j$ and assuming a strongly simplified, uniform electron wavefunction $\psi(\bar{r})=\sqrt{2/N\nu_0}$ over the involved nuclei and where $\tilde{Q}=\sum_{j}x_{j}\frac{I^{j}(I^{j}+1)}{S(S+1)}$ and j=In,As,Ga and 

\begin{equation}
\label{eq:Te}
\frac{1}{T_{1e}}=\left(\frac{\tilde{A}}{N\hbar}\right)^2\frac{2f_e\tau_c}{1+(\frac{\delta_z+\delta_n}{\hbar}\tau_c)^2}
\end{equation}

The fraction of time the QD contains an electron is $f_e$ and the electron Zeeman splitting due to the external magnetic field is $\delta_z$, where $\delta_z^e=\delta_z+\delta_n$. We have assumed for simplicity that $T_d$ is an average nuclear decay constant, independent of the nuclear species. 

Equation \ref{eq:Model} has only one real solution when $\delta_z=g_e\mu_B B_{ext}$ and $\delta_n$ have the same sign, but may have up to three solutions when the signs are opposite, depending on the experimental conditions that determine $\tau_c$. 
This accounts for the existence of bistability effects, which is a general feature of the coupled electron-nuclear spin system, observed in semiconductor bulk, quantum wells and dots \cite{Binet97,Kal92,San2004,braun2006}. 
 
Figure 3 shows that the bistabiliy disappears at higher temperature and $\tau_c$ does indeed decrease with temperature, which seems to confirm our hypothesis. As the dots are separated from the p-doped layer by a tunnel barrier, one could argue that the tunnelling events become more frequent at higher temperature as the energy levels involved get thermally broadened \cite{bu2004} and co-tunneling rates increase \cite{Smith2005,Dreiser2007}. The correlation time $\tau_c$ does not only represent the time the electron is in the dot to interact with the nuclear spins via the hyperfine interaction, it can also be interpreted as the time the nuclei interact with an electron of well defined spin \cite{Abra}. $\tau_c$ will therefore depend on the electron spin relaxation time, which does get shorter at higher lattice temperature \cite{BraunUn}. This is in agreement with the evolution between from figure 3a to 3c: The maximum average spin of the optically injected electron decreases from $S_e^z=0.38$ to $0.2$ when going from 5 to 45K. The values of $\delta_n$ plotted as a function of temperature in figure 1 correspond to the maximum electron polarization at the respective temperature. In table 1 $\delta_n$ is compared for different temperatures for a constant electron spin of $S_e^z=0.2$, showing a 50\% increase of $\delta_n$ when going from 4 to 45K. 

\begin{table}
\caption{\label{tab:table2} Comparison of the OHS $\delta_n$ for $\sigma-$ excitation for the same dot as in figure 1 and 3 for a measured circular polarization degree of $P_c=-40\%$ corresponding to $S_e^z=0.2$ for different  sample temperatures}
\begin{ruledtabular}
\begin{tabular}{ccccc}  % here I choose if it is centred or left etc
Temperature [K] & 4 & 20 & 30 & 45 \\
\hline
 \\
 $\delta_n [\mu eV]$ & 68 $\pm 10$ & 69 $\pm 10$ & 70 $\pm 10$ & 102 $\pm 10$ \\

\end{tabular}
\end{ruledtabular}
\end{table}

It is interesting to note that the effective, internal magnetic field $B_n$ can completely cancel the applied field $B_{ext}$. In this case the electron is exposed to a near zero total field, and one could suspect that electron spin depolarization due to the fluctuating hyperfine field becomes important \cite{pif05}. It can be shown that this is not the case as average time for a flip-flop to take place is always much longer than the radiative lifetime of $X^+$.

The fits in figure 3 have been achieved by varying not only $\tau_c$, but in addition the nuclear spin relaxation time $T_d$. 
Varying either $\tau_c$ or $T_d$ in equation \ref{eq:Model} does not reproduce the experiment, as can be seen by the additional curves in figure 3(c). $T_d$ and $\tau_c$ fit two distinct characteristics of the $\delta_n$ cycle. Adjusting $\tau_c$ allows to fit the width of the bistability region, and in the absence of a bistability, the curvature close to the inflection point of the cycle. $T_d$ determines the maximum nuclear polarization that can be created and fits therefore the extremes of the $\delta_n$ cycle. Our data indicate that the nuclear spin relaxation time $T_d$ gets considerably longer as $\tau_c$ gets shorter. Our result is surprising as both the electron and nuclear spin relaxations time decrease at higher temperature, when measured independently in PL and NMR experiments, respectively. But the situation here is different, as we are investigating the nuclear spin relaxation time of the nuclei in a QD that are coupled to the spin of an electron in the lowest conduction state. This suggests that the presence of the electron in the dot is contributing to the nuclear spin relaxation \cite{Malet2007}. One hypothesis is to link the presence of the $X^+$ to a  local distortion of the lattice, which gives rise to a modification of local electric field gradients. This would amplify nuclear spin relaxation via the modulation of quadropular effects. When the correlation time is reduced, this particular nuclear spin relaxation mechanism might become less efficient. This would explain why $T_d$ increase as $\tau_c$ decreases with temperature. Another nuclear depolarization mechanism could be due to the indirect coupling of nuclear spins in the QD due to the presence of a conduction electron during the radiative lifetime of $X^+$  \cite{Abra,Malet2007,Klauser04}. We have carried out similar experiments to the ones presented in figures 1-3 on the negatively charged excitons $X^-$ in an n-Schottky structure. Dynamical polarization mechanisms for $X^-$ and $X^+$ at low temperature are compared in Ref. \cite{Eble}. Also for $X^-$ the bistability disappears at higher temperature, which can be explained by a shortening of $\tau_c$ as in the $X^+$  case. But in contrast to the $X^+$, the OHS for the $X^-$ does not increase with temperature. The best fits of the $X^-$ data were achieved with a constant nuclear spin relaxation time $T_d$ \cite{xm}. Nuclear depolarization mechanisms linked to the presence of electrons in the dot will be more effective in the $X^-$ case, where an electron is left behind following radiative recombination \cite{Malet2007}. Comparing the $X^+$ and the $X^-$ case indicates that in order to polarize a maximum number of nuclei at higher temperature a shorter $\tau_c$ and simultaneously a longer $T_d$ are necessary. To find the exact dependence of $T_d$ on temperature a more direct measurement would be desirable \cite{Malet2007}.  

In conclusion, the degree of nuclear polarization that can be achieved in a single QD through optical pumping of the $X^+$ exciton depends on the equilibrium that can be established between the flip-flop process of the electron and nuclear spin at the origin of the nuclear polarization on the one hand, and nuclear spin relaxation mechanisms on the other hand.  We have shown that by increasing the sample temperature we achieve a higher nuclear polarization as the nuclear spin generation is \emph{more} and the nuclear spin relaxation is \emph{less} efficient. The flip-flop process does become more likely with increasing temperature as the spin-up and spin-down levels broaden. This is due to a shorter correlation time $\tau_c$ at higher temperature. The exact temperature for which $\delta_n(\sigma^+)$ starts to increase significantly varies from dot to dot in the sample. This is due to the different electron g-factors that result in different Zeeman splittings as well as different strain fields and dot compositions determining $\delta_n$.

\textbf{Acknowledgements:} We thank Daniel Paget for fuitful discussions and ANR MOMES and the IUF for financial support . 
\\
\bibliographystyle{apsrev}

\begin{thebibliography}{36}
\bibitem[{Bra()}]{BraunUn}
\bibinfo{note}{Measurements performed on p doped dot ensembles show a decrease
  of the spin relaxation time from about 10ns at 10K to 1.5ns at 50K (not shown
  here), in agreement with the observed average spin polarization}.
\bibitem[{xm()}]{xm}
\bibinfo{note}{A detailed analysis of the $X^-$ case will be published
  elsewhere.}
\end{thebibliography}

\end{document}